\documentclass[a4paper,11pt]{article}
\usepackage{pos}

\usepackage{slashed}
\usepackage{dsfont}
\usepackage{amsmath}
\usepackage{graphicx}
\usepackage{subcaption}
\usepackage{xspace}

\newcommand{\I}{\mathds{1}}
\newcommand{\Z}{\mathds{Z}}
\newcommand{\Tr}{\mathrm{Tr}}
\newcommand{\DCSB}{D$\chi$SB\xspace}
\newcommand{\muhat}{\hat{\mu}}
\newcommand{\nuhat}{\hat{\nu}}
\newcommand{\munu}{{\mu\nu}}

\title{Impact of centre vortex removal on the Landau-gauge quark propagator in dynamical QCD}
\ShortTitle{Centre vortex removal and the Landau-gauge quark propagator in dynamical QCD}

\author*{Adam Virgili}
\author{Waseem Kamleh}
\author{Derek Leinweber}

\affiliation{Special Research Centre for the Subatomic Structure of Matter, \\
                School of Physical Sciences, University of Adelaide, South Australia 5005, Australia}

\emailAdd{adam.virgili@adelaide.edu.au}
\emailAdd{waseem.kamleh@adelaide.edu.au}
\emailAdd{derek.leinweber@adelaide.edu.au}

\abstract{The important role of centre vortices in dynamical chiral symmetry breaking and corresponding dynamical mass generation has been demonstrated in pure gauge studies of the Landau-gauge quark propagator. 
We present initial results of our investigation into the impact of centre vortex removal on the Landau-gauge propagator on dynamical gauge fields. 
To ensure sensitivity to the topology of the gauge fields, the propagator is computed with overlap fermions.
Upon removal of centre vortices we find that dynamical mass generation is suppressed.
The quark renormalisation function remains remarkably flat except in the deep infrared where it exhibits significant suppression. 
These effects become more prominent at lighter quark masses.} 

\FullConference{%
 The 38th International Symposium on Lattice Field Theory, LATTICE2021
  26th-30th July, 2021
  Zoom/Gather@Massachusetts Institute of Technology
}

\begin{document}
\maketitle

\section{Centre vortices}

Low-energy, nonperturbative QCD is dominated by two main features -- dynamical chiral symmetry breaking (\DCSB) and confinement. 
It is thought these features are mediated by the topological structure of the QCD vacuum. 
Candidates include Abelian monopoles, instantons, and centre vortices.
The latter is the focus of this work.

Centre vortices are revealed through projecting each link to an element of the centre $\mathrm{Z}(N)$ of $\mathrm{SU}(N)$, where
\begin{equation}
    \mathrm{Z}(N) = \{ e^{i 2 \pi n / N} \I\, \vert\, n \in \Z_N \}
\end{equation}
is the set of elements in $\mathrm{SU}(N)$ which commute with every other element of the group.
To obtain the centre-projected links, the gauge field is fixed to maximal centre gauge (MCG) by choosing the gauge transform $U_\mu(x) \to U_\mu^\Omega(x)$ which maximises 
\begin{equation}
    \sum_{x,\mu}\left|{\Tr \, U_\mu^\Omega(x)} \right|^2 \, .
\end{equation}
Each link is then projected to an element of $\mathrm{Z}(3)$, such that
\begin{equation}
    U_\mu^\Omega(x) \to Z_\mu(x) = e^{i\frac{2\pi}{3}n_\mu(x)}\I
\end{equation}
where
\begin{equation}
    n_\mu(x) =
    \begin{cases}
        0 , \text{ if } \arg\Tr\,U^\Omega_\mu(x) \in \left(-\frac{\pi}{3}\,,+\frac{\pi}{3}\right) \,, \\
        1 , \text{ if } \arg\Tr\,U^\Omega_\mu(x) \in \left(+\frac{\pi}{3}\,,+\pi\right) \,, \\
        -1 , \text{ if } \arg\Tr\,U^\Omega_\mu(x) \in \left(-\pi\,,-\frac{\pi}{3}\right) \,.
    \end{cases}
\end{equation}
Centre vortices are identified by the vortex flux through each plaquette $P_{\munu}(x)$, where
\begin{equation}
    P_{\mu\nu}(x) = Z_\mu(x)\,Z_\nu(x+\muhat)\,Z_\mu^\dagger(x+\nuhat)\,Z_\nu^\dagger(x) = e^{i\frac{2\pi}{3}p_\munu(x)}\I \, ,
\end{equation}
corresponds to a vortex flux value $p_\munu(x) \in \left\{-1,0,1\right\}$. 
A plaquette with vortex flux $p_\munu(x)=\pm1$ is identified as being pierced by a vortex.

A vortex-removed link $U^{\mathrm{VR}}_\mu(x)$ is simply the product of the MCG-fixed link and the inverse of its centre-projected link, given by
\begin{equation}
    U^{\mathrm{VR}}_\mu(x) = Z_\mu^\dagger(x) \, U_\mu^\Omega(x)\,.
\end{equation}

\section{Centre vortices and dynamical chiral symmetry breaking}

This work investigates the impact of centre-vortex removal on the overlap Landau-gauge quark propagator.
It is important to elucidate why this is of interest. 
The overlap Landau-gauge quark propagator $S(p)$ can be written in the form
\begin{equation}
    S(p) = \frac{Z(p)}{i\slashed{q} + M(p)}\, ,
    \label{eq:qp}
\end{equation}
where $Z(p)$ is  the renormalisation function and $M(p)$ is the mass function.
The infrared behaviour of the mass function, specifically, the presence of dynamical mass generation, provides a clear signal of \DCSB.
This is thought to be mediated by the topological structure of the QCD vacuum. 
One such candidate is centre vortices.

There is a strong body of evidence to suggest that centre vortices underpin \DCSB in QCD.
It is well-established that centre vortices underpin \DCSB in $\mathrm{SU}(2)$ gauge theory~\cite{deForcrand:1999our,Engelhardt:2002qs,Bornyakov:2007fz,Hollwieser:2008tq,Bowman:2008qd,Hollwieser:2013xja,Hollwieser:2014soz}.
Further, a number of studies in pure $\mathrm{SU}(3)$ gauge theory lend credence to the important role of centre vortices underpinning \DCSB in QCD.
Hadron spectra computed with Wilson fermions on vortex-removed backgrounds display an absence of \DCSB~\cite{OMalley:2011aa}.
Furthermore, a similar study employing overlap fermions was not only able to demonstrate the restoration of chiral symmetry on a vortex-removed background, but was able to reproduce the salient features of the spectrum on a vortex-only background~\cite{Trewartha:2017ive}.
Whilst a study of the AsqTad Landau-gauge quark propagator found weak dependence on centre vortices~\cite{Bowman:2010zr}, this was contradicted by a subsequent study which employed overlap fermions~\cite{Trewartha:2015nna}. 
This study is discussed in detail in the next section.
Combined, these studies present a strong case for centre vortices as the primary mediator of \DCSB in pure $\mathrm{SU}(3)$ gauge theory. 

\section{Overlap Landau-gauge quark propagator in pure SU(3) gauge theory}
\label{sec:qppg}

A previous study~\cite{Trewartha:2015nna} of the Landau-gauge quark propagator in pure $\mathrm{SU}(3)$ gauge theory employed overlap fermions on $20^3 \times 40$ Luscher-Weisz $\mathcal{O}(a^2)$ mean-field improved action ensembles with lattice spacing $a=0.125$ fm.
The mass $M(p)$ and renormalisation $Z(p)$ functions were computed on ensembles comprised of the original (untouched) gauge fields, as well as vortex-removed and vortex-only ensembles, respectively.

\begin{figure}[t]
    \centering
    \begin{subfigure}[b]{0.49\textwidth}
        \centering
        \includegraphics[width=\linewidth]{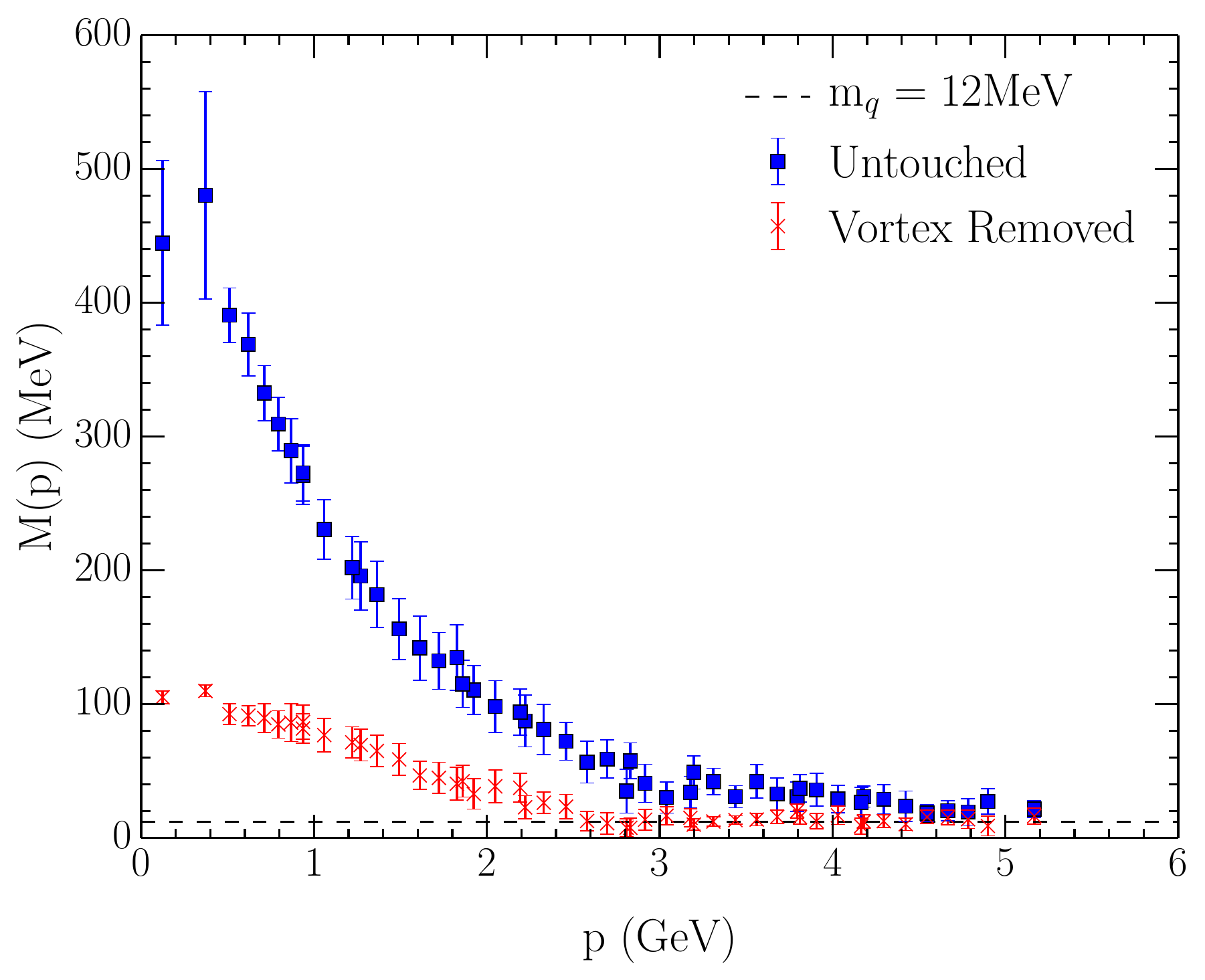}
    \end{subfigure}
    \begin{subfigure}[b]{0.49\textwidth}
        \centering
        \includegraphics[width=\linewidth]{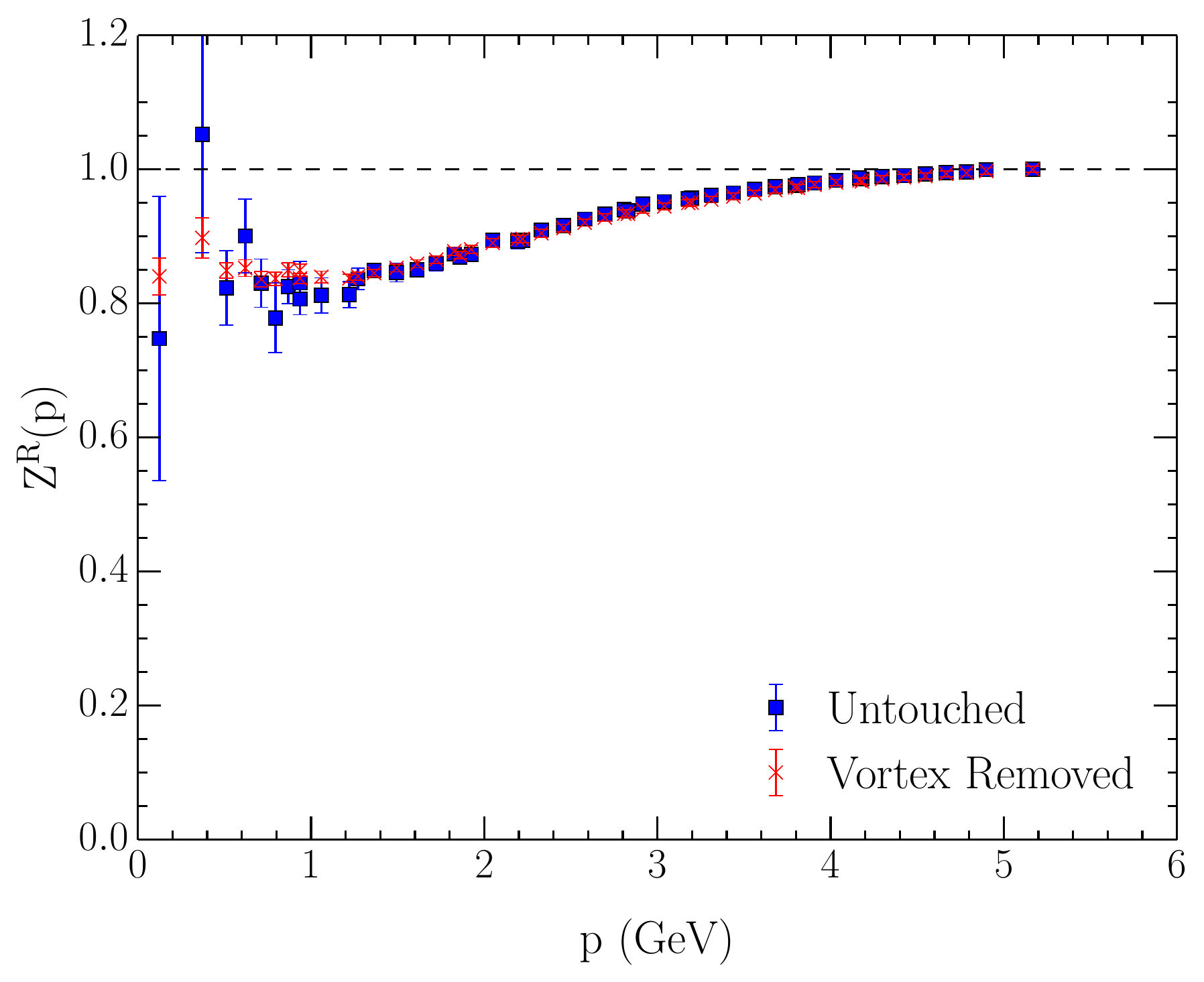}
    \end{subfigure}
    \caption{The mass $M(p)$ (left) and renormalisation $Z(p)$ (right) functions on untouched (blue) and vortex-removed (red) backgrounds in pure gauge theory. Reproduced from Figure 1 (a) and (b) of Ref.~\cite{Trewartha:2015nna}.}
    \label{fig:PGUTVR}
\end{figure}

The functions computed on untouched and vortex-removed ensembles at the lightest quark mass considered, $m_q=12$ MeV, are compared in Figure~\ref{fig:PGUTVR}.
Whilst the respective renormalisation functions are largely similar, we see a clear suppression of dynamical mass generation in the vortex-removed mass function. 
As dynamical mass generation is a clear signal of \DCSB, its suppression upon vortex removal is strong evidence of the important role of centre vortices in \DCSB.

\begin{figure}[t]
    \centering
    \begin{subfigure}[b]{0.49\textwidth}
        \centering
        \includegraphics[width=\linewidth]{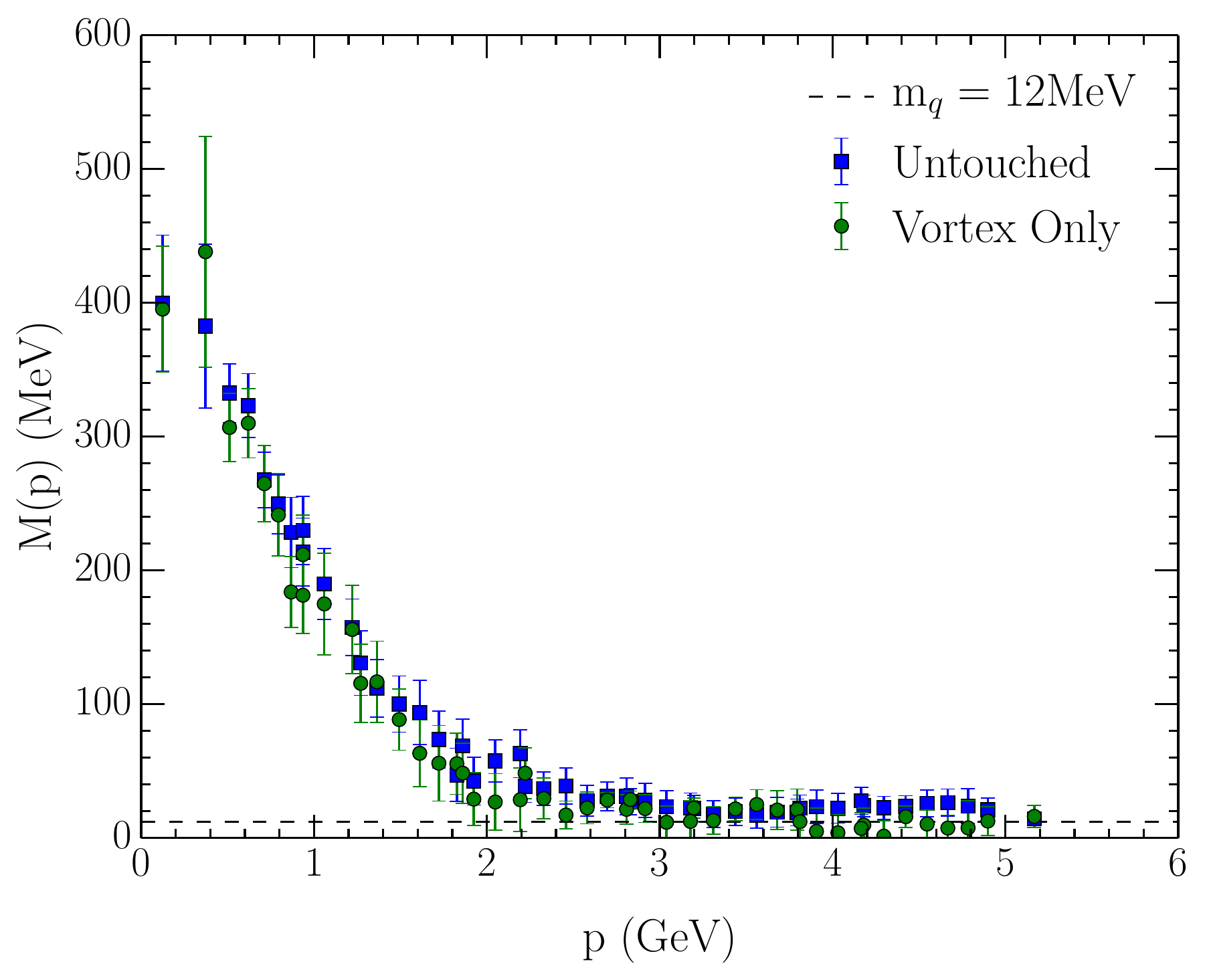}
    \end{subfigure}
    \begin{subfigure}[b]{0.49\textwidth}
        \centering
        \includegraphics[width=\linewidth]{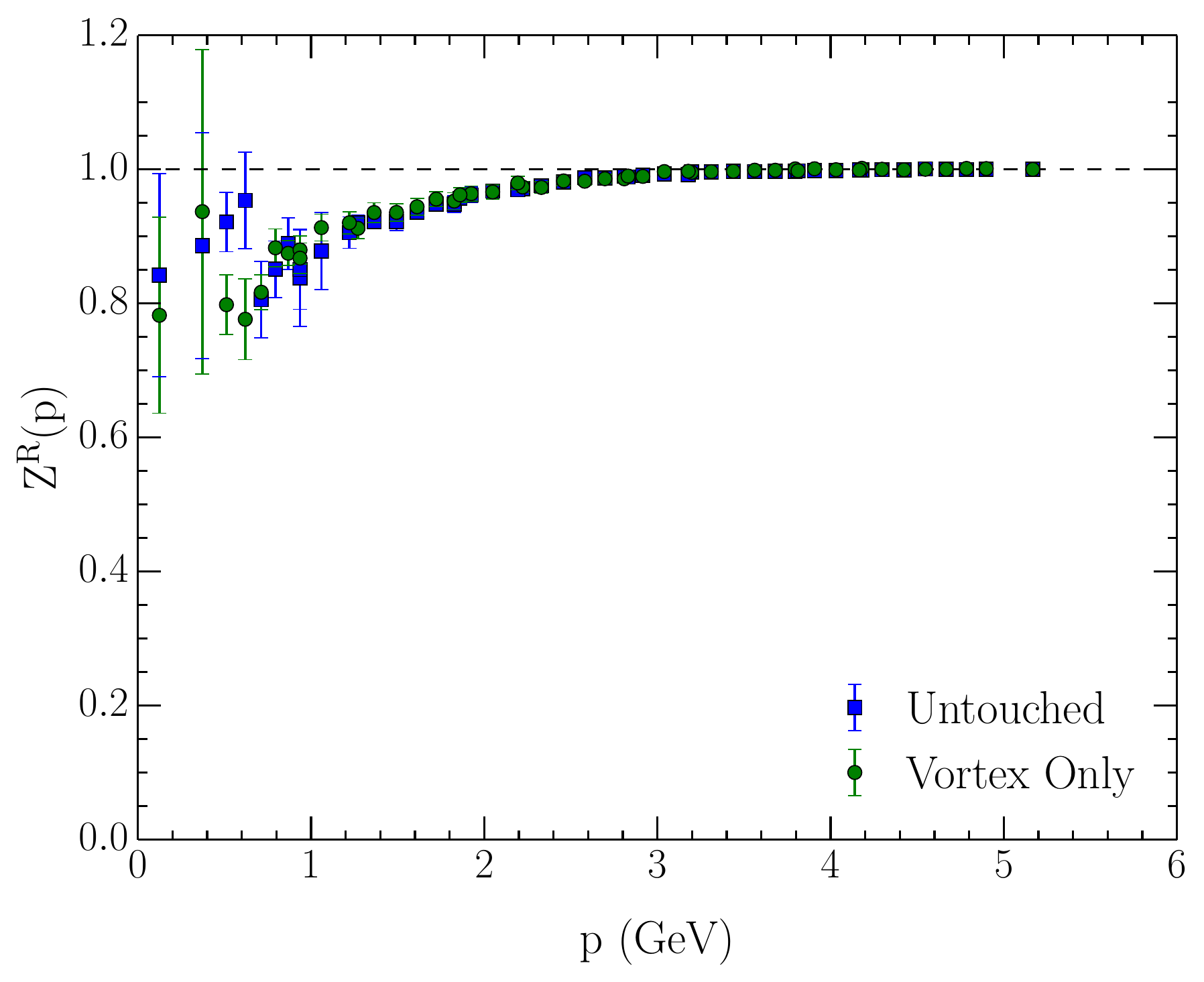}
    \end{subfigure}
    \caption{The mass $M(p)$ (left) and renormalisation $Z(p)$ (right) functions on untouched (blue) and vortex-only (green) backgrounds in pure gauge theory. Reproduced from Figures 3 (a) and (b) of Ref.~\cite{Trewartha:2015nna}.}
    \label{fig:PGUTVO}
\end{figure}

A similar comparison was performed between the untouched and vortex-only ensembles. 
For this comparison, both ensembles underwent 10 sweeps of $\mathcal{O}(a^4)$-improved cooling, necessary for the vortex-only gauge fields to satisfy the smoothness condition of the overlap Dirac operator. 
Results are presented in Figure~\ref{fig:PGUTVO}.
Once again, we see that the renormalisation functions are largely similar.
However, the untouched mass function, and most significantly dynamical mass generation, is reproduced on the vortex-only background.

The combination of these results -- the suppression of dynamical mass generation upon vortex removal, and its reproducibility on a vortex-only background -- present a compelling case for centre vortices as the primary mediator of \DCSB in pure $\mathrm{SU}(3)$ gauge theory.
This raises a pertinent question.
Does this role for centre vortices in \DCSB extend to full, dynamical QCD?
This is the fundamental question this work seeks to answer.

\section{Overlap Landau-gauge quark propagator in dynamical QCD}

\subsection{Simulation parameters}
The overlap Landau-gauge quark propagator was computed on untouched and vortex-removed $32^3 \times 64$ PACS-CS 2+1 flavour ensembles ~\cite{PACS-CS:2008bkb}, each containing 26 configurations at $m_\pi = 156$ MeV.
The fat link irrelevant clover (FLIC) overlap fermion action~\cite{Zanotti:2001yb,Kamleh:2001ff} was employed at seven quark masses $m_q = 6,\,8,\,9,\,19,\,28,\,56,\,84$ MeV where the lightest mass was tuned to match the pion mass of the ensemble.

\subsection{Initial results}
\begin{figure}[t]
    \centering
    \begin{subfigure}[b]{0.49\textwidth}
        \centering
        \includegraphics[width=\linewidth]{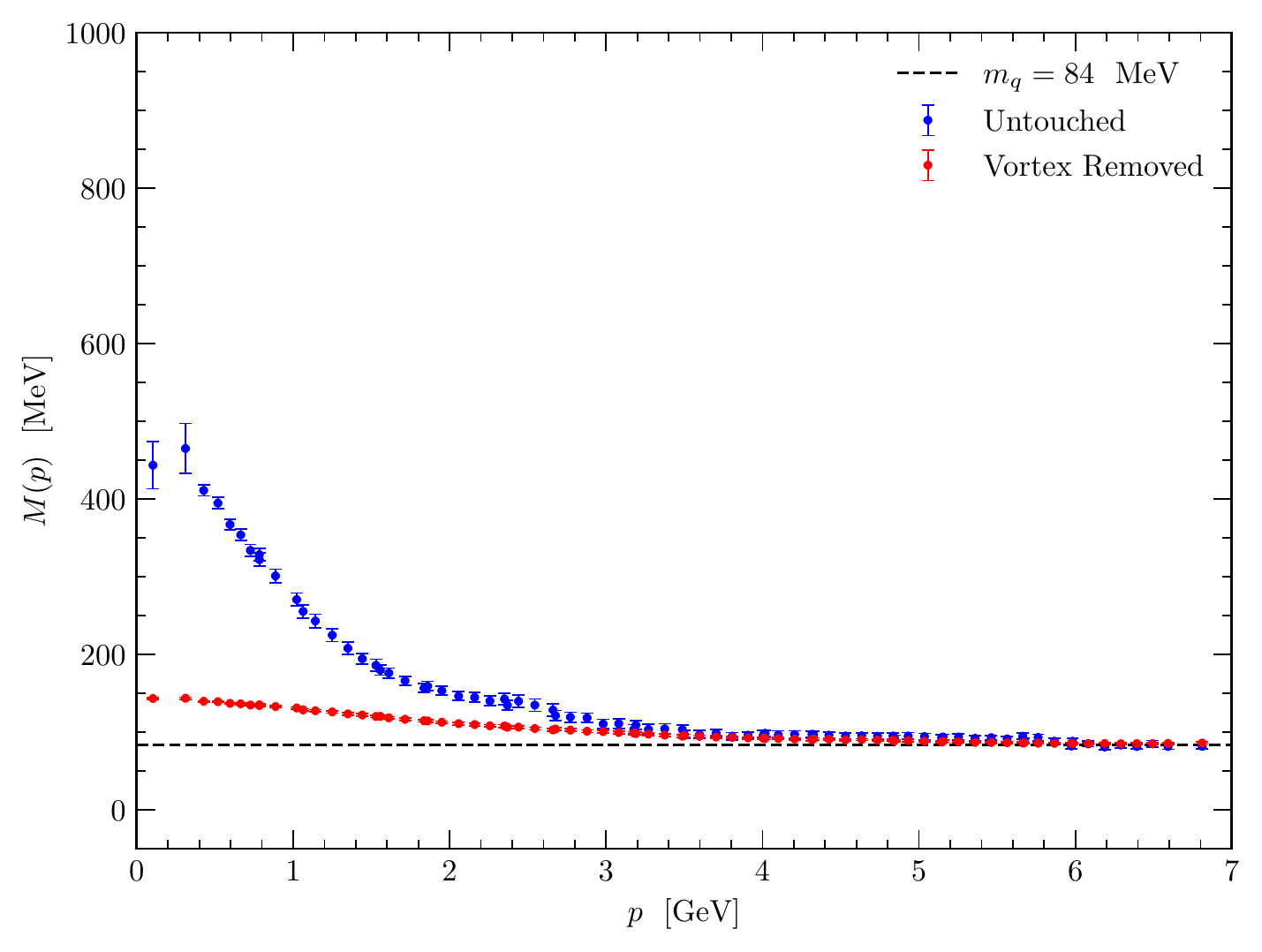}
    \end{subfigure}
    \begin{subfigure}[b]{0.49\textwidth}
        \centering
        \includegraphics[width=\linewidth]{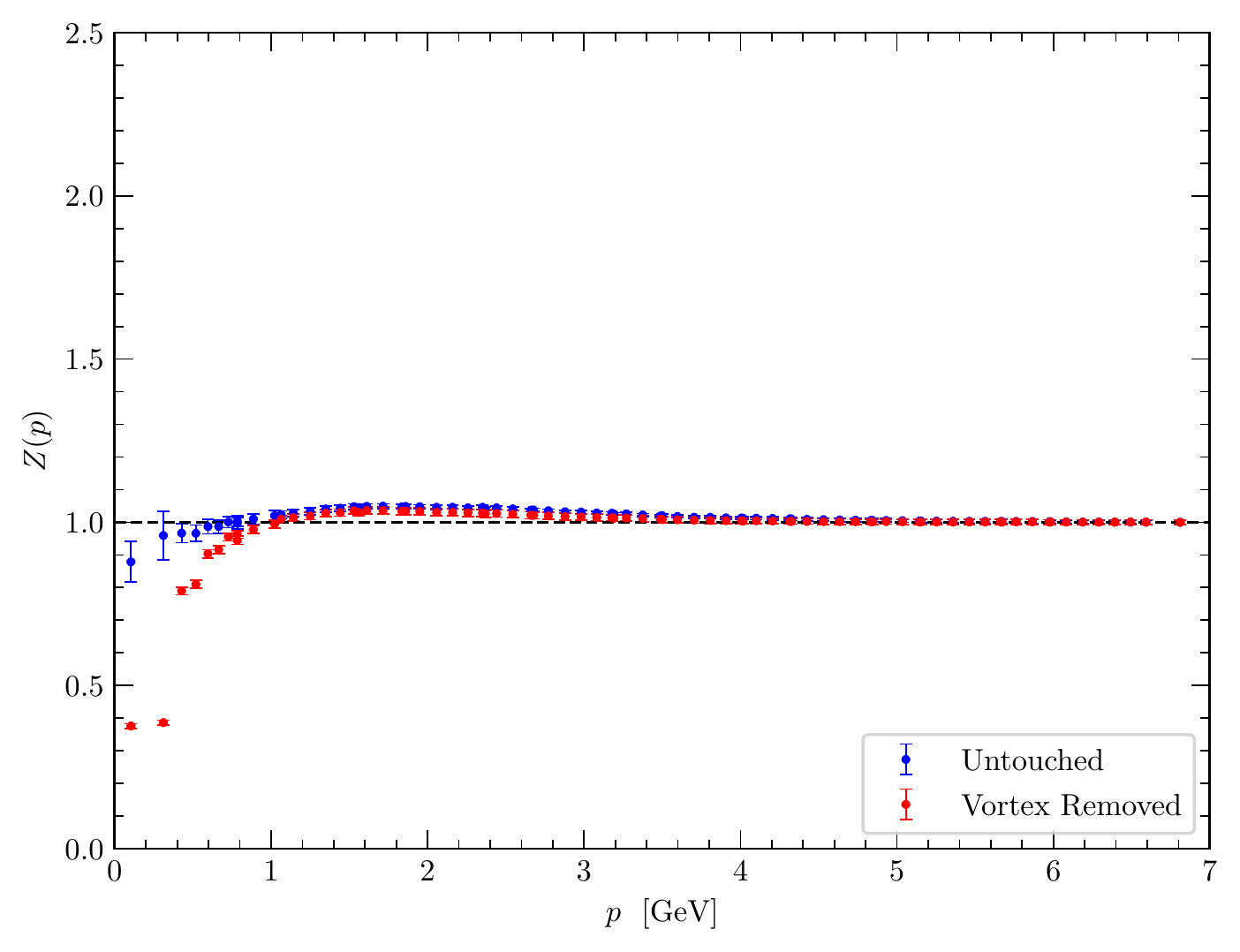}
    \end{subfigure}
    \begin{subfigure}[b]{0.49\textwidth}
        \centering
        \includegraphics[width=\linewidth]{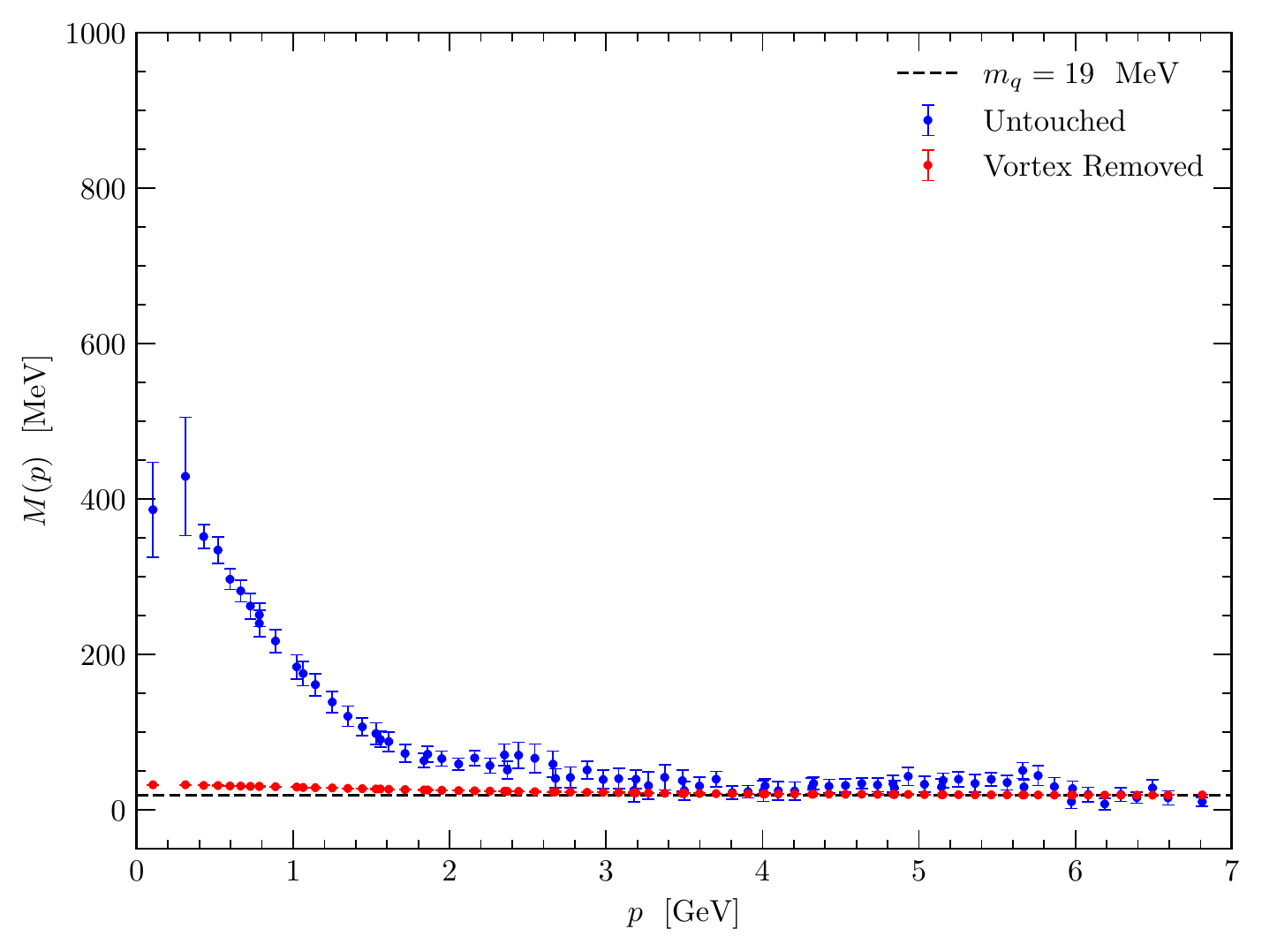}
    \end{subfigure}
    \begin{subfigure}[b]{0.49\textwidth}
        \centering
        \includegraphics[width=\linewidth]{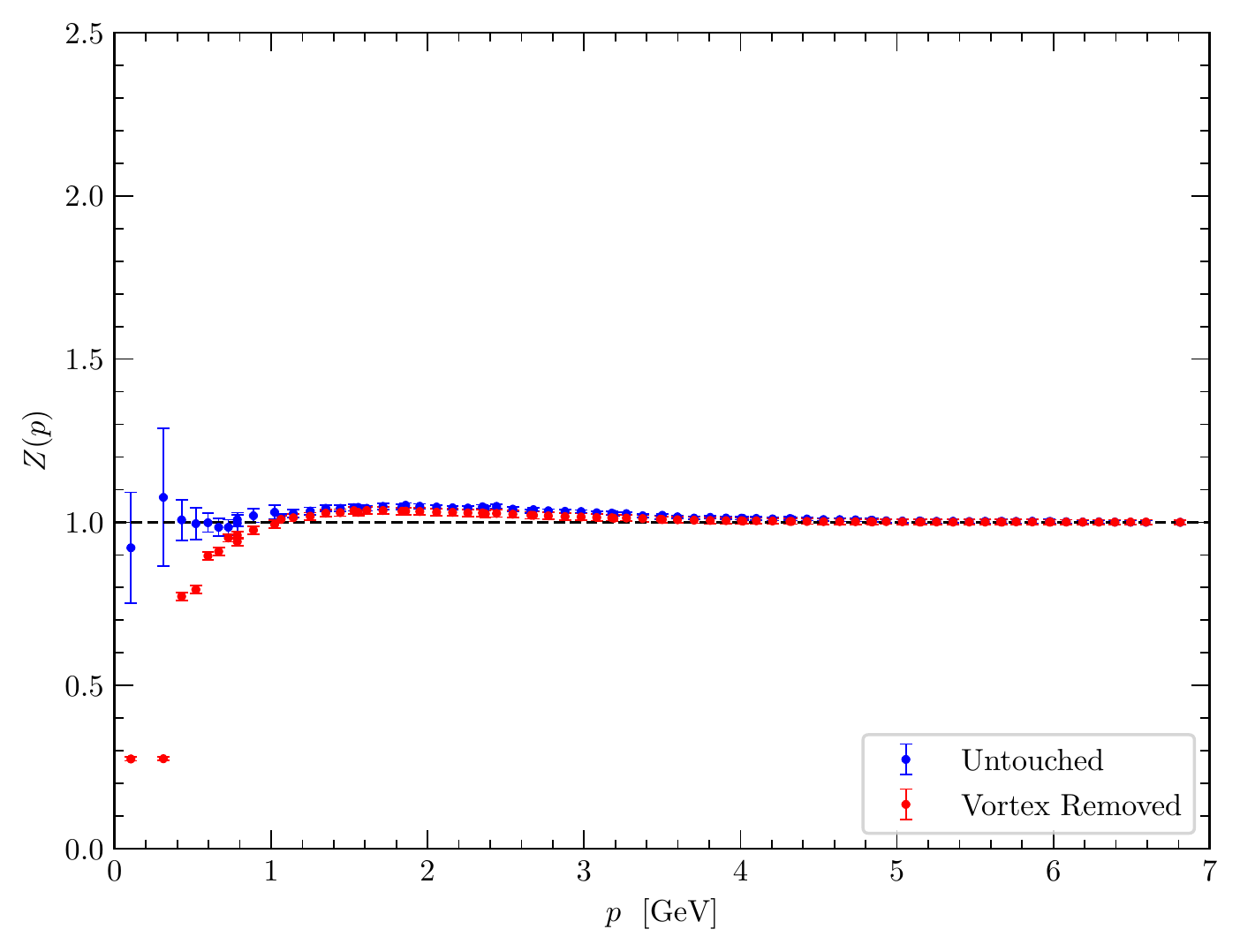}
    \end{subfigure}
    \caption{The mass $M(p)$ (left) and renormalisation $Z(p)$ (right) functions on untouched (blue) and vortex-removed (red) backgrounds for $m_q=84$ MeV (top), and $m_q=19$ MeV (bottom).}
    \label{fig:QCDUTVR}
\end{figure}

The respective untouched and vortex-removed mass $M(p)$ and renormalisation $Z(p)$ functions for the middle ($m_q=19$ MeV) and heaviest ($m_q=84$ MeV) quark masses are presented in Figure~\ref{fig:QCDUTVR}.
At the heaviest mass, dynamical mass generation in the mass function is suppressed upon vortex removal, consistent with pure gauge theory results.
On the other hand, in contrast to pure gauge theory, the renormalisation function is remarkably flat but suffers significant suppression in the deep infrared upon vortex removal.
This suppression is even more significant at the middle mass.
Meanwhile, only a remnant of dynamical mass generation persists at the middle mass upon vortex removal.
This seems to suggest that the degree of dynamical mass generation which persists in the mass function after vortex removal reduces as the quark mass, governing explicit chiral symmetry, moves closer to the dynamical point.
The behaviour of the renormalisation function can also be understood within this hypothesis. 
Recall from Equation~(\ref{eq:qp}) that the overlap Landau-gauge quark propagator can be written as
\begin{equation}
    S(p) = \frac{Z(p)}{i\slashed{q} + M(p)}\, .
    \label{eq:exp1}
\end{equation}
If dynamical mass generation vanishes upon vortex removal as $m_q \to 0$, then 
\begin{equation}
    M(p) \approx 0 \ \ \forall \, p \,.
    \label{eq:exp2}
\end{equation}
Given that 
\begin{equation}
    q \to 0 \text{ as } p \to 0 \, , 
    \label{eq:exp3}
\end{equation}
it must be that
\begin{equation}
    Z(p) \to 0 \text{ as } p \to 0 \, ,
    \label{eq:exp4}
\end{equation}
in a finite system as $S(p)$ remains finite.
This not only explains the presence of significant infrared suppression upon vortex removal, but also its apparent increasing significance as the quark mass moves closer to the dynamical point.

\section{Summary and future work}

\subsection{Summary}

\DCSB is a main feature of low-energy, nonperturbative QCD.
There is significant evidence from existing studies in pure $\mathrm{SU}(3)$ gauge theory which suggest that centre vortices are the primary mediator of \DCSB.
Through the study of the overlap Landau-gauge quark propagator this work investigates the extent to which centre vortices underpin \DCSB in dynamical QCD.
The mass and renormalisation functions of the dynamical quark propagator have been computed on respective untouched and vortex-removed ensembles, and compared.
At the heaviest mass, persistence of dynamical mass generation upon vortex removal is consistent with existing pure gauge theory results.
However, dynamical mass generation is diminished at lighter quark masses as explicit chiral symmetry breaking is reduced.
Meanwhile, the renormalisation function is remarkably flat, but suffers significant infrared suppression upon vortex removal, with greater significance at lighter quark masses.
Understood together, these results are suggestive of vanishing dynamical mass generation towards the physical point.
Nevertheless, present results only lend further credibility to the notion of centre vortices as the primary mediator of \DCSB in dynamical QCD.

\subsection{Ongoing and future work}

Results for the quark propagator at the dynamical point are currently in preparation.
Once these are obtained, the next step is to compute and compare the quark propagator on respective, equivalently smeared, untouched and vortex-only ensembles.
As alluded to in Section~\ref{sec:qppg}, it is necessary to smooth vortex-only gauge fields to satisfy the smoothness condition of the overlap Dirac operator.
In past pure gauge theory studies, cooling algorithms have sufficed to this end.
Such algorithms, however, are not ideal for dynamical gauge fields.
An ideal smoothing algorithm would not only be analytical, but preserve a memory of the underlying vortex structure.
What such an algorithm looks like, or whether one exists, merits investigation.

\acknowledgments{We thank the PACS-CS Collaboration for making their 2+1 flavour configurations available via the International Lattice Data Grid (ILDG). This research was undertaken with the assistance of resources provided by the Pawsey Supercomputing Centre with funding from the Australian Government and the Government of Western Australia; and  from the National Computational Infrastructure (NCI), provided through the National Computational Merit Allocation Scheme and supported by the Australian Government through Grant No. LE190100021 via the University of Adelaide Partner Share. This research is supported by the Australian Research Council through Grants No. DP190102215 and DP210103706.
WK is supported by the Pawsey Supercomputing Centre through the Pawsey Centre for Extreme Scale Readiness (PaCER) program.}

\bibliographystyle{JHEP}
\bibliography{Bibliography}

\end{document}